\documentclass[10pt]{article}
\usepackage[OE]{express}
\usepackage{graphicx}
\usepackage{dcolumn}
\usepackage{bm}
\usepackage{array}
\usepackage{hyperref}
\usepackage{amsmath}
\usepackage{multirow}
\usepackage{cite}

\usepackage{mathrsfs}
\usepackage{}
\usepackage{makecell}
\usepackage{multirow}
\usepackage{dcolumn}

\newlength{\figwidth} 
\newcommand{\beq}{\begin{equation}}

\newcommand{\beql}[1]{\begin{equation}\label{#1}}
\newcommand{\eeq}{\end{equation}}
\newcommand{\bsp}{\begin{split}}
\newcommand{\esp}{\end{split}}

\newcommand{\Eq}[1]{Eq.~(\ref{#1})}
\newcommand{\Equation}[1]{Equation~(\ref{#1})}

\newcommand{\Fig}[1]{Fig.~\ref{#1}}
\newcommand{\Figs}[1]{Figs.~\ref{#1}}

\newcommand{\Table}[1]{Table.~(\ref{#1})}
\begin{document}
\title{Classification of symmetry properties of waveguide modes in presence of gain/losses, anisotropy/bianisotropy, or continuous/discrete rotational symmetry}
\author{Zhongfei Xiong\authormark{1},  Weijin Chen\authormark{1}, Peng Wang\authormark{1}, Yuntian Chen\authormark{1,2,*}}
\address{\authormark{1}School of Optical and Electronic Information, Huazhong University of Science and Technology, Wuhan, 430074, China\\
\authormark{2} Wuhan National Laboratory of Optoelectronics, Huazhong University of Science and Technology, Wuhan, China.}  
\email{\authormark{*}yuntian@hust.edu.cn}

\begin{abstract}
We study the symmetric properties of waveguide modes in presence of gain/losses, anisotropy/bianisotropy, or continuous/discrete rotational symmetry. We provide a comprehensive approach to identity the modal symmetry by constructing a $4\times4$ waveguide Hamiltonian and searching the symmetric operation in association with the corresponding waveguides. We classify the chiral/time reversal/parity/parity time/rotational symmetry for different  waveguides, and provide the  criterion for  the aforementioned  symmetry operations. Lastly, we provide   examples  to illustrate how the symmetry operations can be used to classify the modal properties from the symmetric relation between modal profiles of several different waveguides.
\end{abstract}


\section{Introduction}

It is well-accepted that  there are beautiful symmetric structures  embedded in Maxwell's equations, i.e., the dual symmetry between electric and magnetic fields,  time reversal symmetry, and many others as explained in \cite{Fushchich1987}. Those symmetries on one hand could be used to simplify our understanding of mode hybridization associated with complicated optical structures \cite{Sakoda}, on the other hand impose certain constraints to electromagnetic response \cite{Alurpl2017}. One also notes that certain optical structures based on combined symmetries of parity and time reversal $\mathcal{PT}$ posses interesting features, i.e., real eigenvalues though the Hamiltonian being non-Hermitian, and exceptional points (EPs) where the transition of  $\mathcal{PT}$ symmetry breaking occurs. It is necessary to study and understand  the  general scenarios where those symmetries can be broken,  leading to astonishing behaviors of light such as  non-reciprocal  or one-way  propagation.   In waveguides, there is an  additional  symmetry, i.e., translation symmetry along the propagation direction. Such translation symmetry  ensures  the modal wave number a constant value, i.e.,  propagation constant $\beta$, which is a typical terminology in waveguides. In analogy of a waveguide mode $\bm E(\bm r)=\bm e(x,y) e^{-i\beta z}$ to a wave-function $\Psi(\bm r,t)=\Psi(\bm r)  e^{-i E t}$ associated  with the stationary  Schr\"odinger equation $H\Psi(\bm r)=E\Psi(\bm r)$, $z$ plays the the role of time t, and $\beta$ plays the role of energy $E$ \cite{SgJohnsonOE2002,ChongPRA2017}. 

In isotropic waveguide, the negative propagating modes (-$\beta$) can be considered as a perfect image of the forward propagating modes ($\beta$). It  is interesting to ask how the forward and backward propagating modes are related, if the waveguide materials contain gain/losses, anisotropy, or bianisotropy?  One also notes that if the waveguide cross-section contains rotational symmetries, the polarization modes associated with the same field configuration may or may not degenerate. Though those results are well documented in the literatures, there is no systematic approach to classify the symmetry properties of waveguide mode using the equivalent Hamiltonian, considering  the analogy of the wave equation of the waveguides with the  stationary  Schr\"odinger equation. Along this line, it is important to point out the waveguide mode are vector fields, in contrast to the scalar wave function  in Schr\"odinger equation.

In this work, we derive the exact Hamiltonian of the waveguide  from Maxwell's equations. In our formulation, we take into account the vectorial nature of electromagnetic  fields in the equivalent waveguide Hamiltonian, which resembles Dirac equation accounting for electrons with positive/negative energies and up/down spin states. By construction, we search for the symmetry operations associated with the Hamiltonian to classify the symmetry properties of waveguide modes in presence of gain/losses, anisotropy/bianisotropy, or continuous/discrete rotational symmetry in the geometric cross-section of the waveguides.

The paper is organized as follows: In Section 2, we outline the construction of Hamiltonian for the waveguide as well as  the description of the symmetry operations. In Section 3, we
apply  the symmetry operations to the  Hamiltonian of different waveguides, and  to classify the symmetric properties of the waveguide modes. Finally, Section 4 concludes the paper.

\section{Theory}
\subsection{Waveguide Hamiltonian}
We take a time harmonic dependence $e^{i \omega t }$ for the electromagnetic waves throughout this paper. The source-free Maxwell's equations for general bianisotropic waveguide read as follows, 
\begin{equation}\label{max1}
\begin{array}{c}
 \left[ {\nabla  \times  +I{k_0}{{\bar{\bm \chi } }_{he}}} \right]{\bm e_{3d}}(x,y,z){\text{ + }}I{k_0}({{\bar {\bm\mu} }_r}){\bm h_{3d}}(x,y,z) = 0, \\
  \left[ {\nabla  \times  - I{k_0}{{\bar{\bm \chi } }_{eh}}} \right]{\bm h_{3d}}(x,y,z) - I{k_0}({{\bar {\bm\varepsilon} }_r}){\bm e_{3d}}(x,y,z) = 0, 
\end{array}
\end{equation} 
where $ {\bm e_{3d}}(x,y,z) = {e_{2d}}(x,y){e^{ - i\beta z}} = \bm e_{2d}^t(x,y){e^{ - i\beta z}} + e_{2d}^z(x,y){e^{ - i\beta z}}$, ${\bm h_{3d}}(x,y,z) = {\bm h_{2d}}(x,y){e^{ - i\beta z}} = \bm h_{2d}^t(x,y){e^{ - i\beta z}} + e_{2d}^z(x,y){e^{ - i\beta z}}$, $\bar{\bm\varepsilon}_r=\begin{pmatrix}
 \bar{\bm\varepsilon}_r^{tt}&\bar{\bm\varepsilon}_r^{tz}\\
\bar{\bm\varepsilon}_r^{zt}&\bar{\bm\varepsilon}_r^{z} \\
\end{pmatrix}=\begin{pmatrix}
 \varepsilon_r^{xx}&\varepsilon_r^{xy}&\varepsilon_r^{xz}\\
\varepsilon_r^{yx}&\varepsilon_r^{yy}&\varepsilon_r^{yz} \\
\varepsilon_r^{zx}&\varepsilon_r^{zy}&\varepsilon_r^{zz}\\
\end{pmatrix}$, $\bar{\bm\mu}_r=\begin{pmatrix}
 \bar{\bm\mu}_r^{tt}&\bar{\bm\mu}_r^{tz}\\
\bar{\bm\mu}_r^{zt}&\bar{\bm\mu}_r^{z} \\
\end{pmatrix}=\begin{pmatrix}
 \mu_r^{xx}&\mu_r^{xy}&\mu_r^{xz}\\
\mu_r^{yx}&\mu_r^{yy}&\mu_r^{yz} \\
\mu_r^{zx}&\mu_r^{zy}&\mu_r^{zz}\\
\end{pmatrix}$, and $\bar{\bm \chi} _{he}=-\bar{\bm \chi} _{eh}^T=i \bar{\bm \chi}=i \left({\begin{array}{*{20}{c}}
  {\chi_{xx}}&{\chi_{xy}} \\ 
  {\chi_{yx}}&{\chi_{yy}}\\ 
  \end{array}} \right)$. \Equation{max1}  can be reformulated into 4 components equation, by eliminating the $e^z_{2d}(x,y)$ and $h^z_{2d}(x,y)$  via the expressions ${e_{2d}^z(x,y) = \frac{{{\nabla _t} \times h_{2d}^t(x,y) - I{k_0}\bar \varepsilon _r^{zt} \cdot e_{2d}^t(x,y)}}{{I{k_0}\bar \varepsilon _r^z}}}$ and ${h_{2d}^z(x,y) =  - \frac{{{\nabla _t} \times e_{2d}^t(x,y) + I{k_0}\bar \mu _r^{zt} \cdot h_{2d}^t(x,y)}}{{I{k_0}\bar \mu _r^z}}}$. The resulted equations for the in-plane field components can be written in a compact form, 
\beq\label{ham}
H \Psi= \beta \Psi,
\eeq 
where the  Hamiltonian $H$ given by 
\begin{tiny}
\[H = \left( {\begin{array}{*{20}{c}}
  { - i{\partial _x}\frac{{\bar \varepsilon _r^{zx}}}{{\bar \varepsilon _r^{zz}}}}-i\frac{\mu_r^{yz}}{\mu_r^{zz}}{\partial_y}+i k_0 \chi_{yx}&{-i{\partial _x}\frac{{\bar \varepsilon _r^{zy}}}{{\bar \varepsilon _r^{zz}}}}+i\frac{\mu_r^{yz}}{\mu_r^{zz}}{\partial_x} +i k_0 \chi_{yy}&{ - {\partial _x}\frac{{{\partial _y}}}{{{k_0}\bar \varepsilon _r^{zz}}}-\frac{k_0}{\mu_r^{zz}}\mu_r^{yz}\mu_r^{zx}  + {k_0}\bar \mu _r^{yx}}&{{\partial _x}\frac{{{\partial _x}}}{{{k_0}\bar \varepsilon _r^{zz}}} -\frac{k_0}{\mu_r^{zz}}\mu_r^{yz}\mu_r^{zy}   + {k_0}\bar \mu _r^{yy}} \\ 
  { - i{\partial _y}\frac{{\bar \varepsilon _r^{zx}}}{{\bar \varepsilon _r^{zz}}}}  +i\frac{\mu_r^{xz}}{\mu_r^{zz}}{\partial_y}  -i k_0 \chi_{xx}&{-i{\partial _y}\frac{{\bar \varepsilon _r^{zy}}}{{\bar \varepsilon _r^{zz}}}} -i\frac{\mu_r^{xz}}{\mu_r^{zz}}{\partial_x} -i k_0 \chi_{xy}&{ - {\partial _y}\frac{{{\partial _y}}}{{{k_0}\bar \varepsilon _r^{zz}}} +\frac{k_0}{\mu_r^{zz}}\mu_r^{xz}\mu_r^{zx}  - {k_0}\bar \mu _r^{xx}}&{{\partial _y}\frac{{{\partial _x}}}{{{k_0}\bar \varepsilon _r^{zz}}} +\frac{k_0}{\mu_r^{zz}}\mu_r^{xz}\mu_r^{zy}  - {k_0}\bar \mu _r^{xy}} \\ 
  {{\partial _x}\frac{{{\partial _y}}}{{{k_0}\bar \mu _r^{zz}}} + \frac{{{k_0}}}{{\bar \varepsilon _r^{zz}}}\bar \varepsilon _r^{yz}\bar \varepsilon _r^{zx} - {k_0}\bar \varepsilon _r^{yx}}&{ - {\partial _x}\frac{{{\partial _x}}}{{{k_0}\bar \mu _r^{zz}}} + \frac{{{k_0}}}{{\bar \varepsilon _r^{zz}}}\bar \varepsilon _r^{yz}\bar \varepsilon _r^{zy} - {k_0}\bar \varepsilon _r^{yy}}&{-i{\partial_x}\frac{\mu_r^{zx}}{\mu_r^{zz}} - i\frac{{\bar \varepsilon _r^{yz}}}{{\bar \varepsilon _r^{zz}}}{\partial _y}}+i k_0 \chi_{xy}&{-i{\partial_x}\frac{\mu_r^{zy}}{\mu_r^{zz}} + i\frac{{\bar \varepsilon _r^{yz}}}{{\bar \varepsilon _r^{zz}}}{\partial _x}}+i k_0 \chi_{yy} \\ 
  {{\partial _y}\frac{{{\partial _y}}}{{{k_0}\bar \mu _r^{zz}}} - \frac{{{k_0}}}{{\bar \varepsilon _r^{zz}}}\bar \varepsilon _r^{xz}\bar \varepsilon _r^{zx} + {k_0}\bar \varepsilon _r^{xx}}&{ - {\partial _y}\frac{{{\partial _x}}}{{{k_0}\bar \mu _r^{zz}}} - \frac{{{k_0}}}{{\bar \varepsilon _r^{zz}}}\bar \varepsilon _r^{xz}\bar \varepsilon _r^{zy} + {k_0}\bar \varepsilon _r^{xy}}&{-i{\partial_y}\frac{\mu_r^{zx}}{\mu_r^{zz}} +i\frac{{\bar \varepsilon _r^{xz}}}{{\bar \varepsilon _r^{zz}}}{\partial _y}}-i k_0 \chi_{xx}&{-i{\partial_x}\frac{\mu_r^{zy}}{\mu_r^{zz}} - i\frac{{\bar \varepsilon _r^{xz}}}{{\bar \varepsilon _r^{zz}}}{\partial _x}}-i k_0 \chi_{yx} 
\end{array}} \right), 
\]
\end{tiny} 
and $\Psi=\left[e_{x}(x,y), e_{y}(x,y),h_{x}(x,y),h_{y}(x,y)\right]^T$ is the eigenstate, which contains the in-plane field components. In  \Eq{ham}, we limit our self to study the mode properties of the waveguide within the truncated mode set, with particular emphasis on the symmetry relations among the polarizations, as well as that between  the forward propagating modes and the backward propagating modes.  The truncated mode set is defined as the waveguide modes, which share the field configuration  labeled by the same quantum numbers in the traverse plane. For simplicity, we investigate the waveguides  with single core structure, the medium of which could be active, lossy, anisotropic or bianisotropic. The geometric cross section of the waveguide core structure could be irregular, or highly symmetric. The background is homogeneous and isotropic. 

Corresponding to the $4\times4$ matrix form Hamiltonian, there will be 4 eigenmodes  $\Psi_1^+$, $\Psi_2^+$, $\Psi_1^-$ and $\Psi_2^-$ in the truncated mode set, with eigenvalue  being $\beta_1^+$, $\beta_2^+$, $\beta_1^-$ and $\beta_2^-$ respectively. The superscript $+$($-$) indicates forward (backward) propagating modes, and we note a pair of orthogonal polarization modes in same direction with subscript 1 or 2. Once  the waveguide  Hamiltonian is known,  the degeneracy of the modes within the truncated mode set can be classified by searching proper symmetry operations. 

In the paper, we concern waveguiding mainly by the refractive index contrast. Thus, the waveguide can be sliced into regions with  piece-wise constant material properties. To perform modal analysis of waveguide,  one finds the eigenfields of each region, and  then apply the boundary condition to connect the fields from different regions such that the eigenfields of the waveguide can be obtained. This procedure shows that the final eigenfield of the waveguide can be seen as certain combination of  the eigenfield of each individual region, though the boundary condition determines how the eigenfields from different region are combined. In any case, the final eigenfield of the waveguide mode obeys the same symmetry as the eigenfield of each individual region, provided the same modal wave number $\beta$ is selected. Thus, the study on  the symmetry properties of the waveguide mode can be reduced to analysis the symmetry properties of  the eigenfied of each individual region, with no need to concern the boundary conditions. In our settings, the background of the waveguide core is air, the symmetry relation of the waveguide mode is essentially determined by the waveguide core, which is our focus in the following sections.

\subsection{Chiral symmetry}
We study the degeneracy between opposite propagating modes ($\Psi_1^+$ and $\Psi_1^-$ or $\Psi_2^+$ and $\Psi_2^-$). Here, an unitary matrix 
\beq\label{sigma}
\sigma  = \left( {\begin{array}{*{20}{c}}
  1&0&0&0 \\ 
  0&1&0&0 \\ 
  0&0&{ - 1}&0 \\ 
  0&0&0&{ - 1} 
\end{array}} \right),
\eeq
is introduced as an operator to describe a chiral transformation.  As the operator $\sigma$ acts on a state $\Psi$, it reverses the sign of transverse magnetic field while the transverse electric fields remain unchanged, and the original and transformed transverse electromagnetic fields  can be seen as left-handed and right-handed systems.  Since the Poynting vector $\bm P$ is defined as $\bm P=\bm{E}\times\bm{H}$, the chiral operation $\sigma$ changes the propagation direction of power flow, thereby builds the connection between forward and backward propagating modes. If the terms $\bar {\bm \varepsilon} _r^{zt}$, $\bar {\bm \varepsilon} _r^{tz}$,$\bar {\bm \mu} _r^{zt}$, $\bar {\bm \mu} _r^{tz}$ and $\bar{\bm \chi}$ in Hamiltonian H  vanish, then H in \Eq{ham} is reduced to,
\beq \label{chiral}
H =\left( {\begin{array}{*{20}{c}}
  0&0&{ - {\partial _x}\frac{{{\partial _y}}}{{{k_0}\bar \varepsilon _r^{zz}}}+ {k_0}\bar \mu _r^{yx}}&{{\partial _x}\frac{{{\partial _x}}}{{{k_0}\bar \varepsilon _r^{zz}}} + {k_0}\bar \mu _r^{yy}} \\ 
  0&0&{ - {\partial _y}\frac{{{\partial _y}}}{{{k_0}\bar \varepsilon _r^{zz}}} - {k_0}\bar \mu _r^{xx}}&{{\partial _y}\frac{{{\partial _x}}}{{{k_0}\bar \varepsilon _r^{zz}}}- {k_0}\bar \mu _r^{xy}} \\ 
  {{\partial _x}\frac{{{\partial _y}}}{{{k_0}\bar \mu _r^{zz}}} - {k_0}\bar \varepsilon _r^{yx}}&{ - {\partial _x}\frac{{{\partial _x}}}{{{k_0}\bar \mu _r^{zz}}} - {k_0}\bar \varepsilon _r^{yy}}&0&0 \\ 
  {{\partial _y}\frac{{{\partial _y}}}{{{k_0}\bar \mu _r^{zz}}} + {k_0}\bar \varepsilon _r^{xx}}&{ - {\partial _y}\frac{{{\partial _x}}}{{{k_0}\bar \mu _r^{zz}}} + {k_0}\bar \varepsilon _r^{xy}}&0&0 
\end{array}} \right).
\eeq 
A close examination shows that  the following relation for the reduced waveguide Hamiltonian H in \Eq{chiral} holds,
\beq\label{chiral symmetry} 
\sigma H{\sigma ^{ - 1}} =  - H,
\eeq 
which means if $\beta_1$ is the eigenvalue of $H$ with eigenstate $\Psi_1$, the $-\beta_1$ would also be the eigenvalue with eigenstate $\sigma\Psi_1$. In other words, for  a given forward propagating  mode, there is  a degenerate backward propagating mode, and the eigen-fields   transform to each other by the symmetry operation $\sigma$, provided that the  constraints on $\bar {\bm \varepsilon} _r$, $\bar {\bm \mu} _r$ and $\bar{\bm \chi}$ are fulfilled. 
\subsection{Time reversal symmetry}

Next, we introduce the time reversal operator $\mathcal{T}:\hat{p}\rightarrow -\hat{p},i\Rightarrow -i$, where $\hat{p}$ is the momentum operator\cite{Ganainy2007_OL,PT}. In general, this operator can be represented as $\mathcal{T}=UK$, where $U$ is a unitary matrix and $K$ is complex conjugation\cite{BernevigTR2013}. The operator $\sigma$ used in chiral symmetry operation is an unitary matrix, and will be used here to replace  $U$, leading to the time reversal operator as follows,
\beq\label{TR} 
\mathcal{T}=\sigma K.
\eeq
As the operator $K$ acts on the Hamiltonian, all the $i$ in \Eq{ham} reverses sign, and all the elements in  the permittivity tensor $\bar{\bm\varepsilon}_r$,  permeability tensor $\bar{\bm\mu}_r$ and $\bar{\bm\chi}$ in \Eq{ham} take the complex
conjugate. If the waveguide is invariant under time reversal operation, which requires all the  these elements in
 material tensors, i.e., $\bar{\bm\varepsilon}_r$,  $\bar{\bm\mu}_r$ and $\bar{\bm\chi}$  to be real numbers, we shall have,
\beq\label{TRsymmetry}
\mathcal{T} H{\mathcal{T}^{ - 1}} = -H
\eeq
Similar to \Eq{chiral}, the Hamiltonian also  reverses sign under the time reversal operation. Therefore,
as a result of \Eq{TRsymmetry}, the forward and backward propagating modes are  degenerated, but up to  a sign difference in the eigenvalues ($\beta$), the  eigenstates are related by operator $\mathcal{T}$. In contrast to  chiral symmetry operator, we don't necessarily need  the  reduced  Hamiltonian in \Eq{chiral} for $\mathcal{T}$ operator, but the time reversal symmetry indeed  requires that  all the elements in the material tensors ($\bar{\bm\varepsilon}_r$,  $\bar{\bm\mu}_r$ and $\bar{\bm\chi}$) to be real. And the transformation between the fields of the degenerate modes, not only needs $\sigma$, but also needs take the complex conjugate. Despite those differences in chiral symmetry operator and the time reversal operator, both can be applied to scenarios, in which $\bar{\bm\chi}$ is zero, $\bar{\bm\varepsilon}_r$,   $\bar{\bm\mu}_r$ are real and without  $tz$,$zt$ elements, and the two symmetry operation yields exactly the same results. Same as chiral symmetry, time reversal operator $\mathcal{T}$ doesn't perform any action on space , thus  there is no constraint on the geometry structure of waveguide.  

\subsection{Parity symmetry}
We proceed to discuss the symmetry operation that changes the coordinates, for example, parity operator $\mathcal{P}$, $\bm r\rightarrow-\bm r$, $\hat{p}\rightarrow -\hat{p}$, where $\bm r$ is the position operator and only contains transverse coordinate $(x,y)$ \cite{Ganainy2007_OL,PT}. The optical properties  of waveguide are essentially determined by  the spatial dependent permittivity and permeability, i.e., $\bar{\bm \varepsilon}_r\left(\bm r \right)$ and $\bar{\bm \mu}_r\left(\bm r \right)$. Considering  $\bar{\bm\chi}$=0, \Eq{ham} can be reformulated as:
\beq\label{equationspace}
H\left(\bm r,\bar{\bm \varepsilon}_r\left(\bm r \right),\bar{\bm \mu}_r\left(\bm r \right)\right)\Psi\left(\bm r\right)=\beta\Psi\left(\bm r\right),
\eeq
the first $\bm r$ in Hamiltonian $H$ represents the coordinates  that get differentiated, all the rest $\bm r$ in \Eq{equationspace} simply represents the spatial dependence of material tensors and wave-function.  The parity operator $\mathcal{P}$ also  contains a unitary matrix $\sigma$ and an operator that reverses coordinate. As the operator $\mathcal{P}$
acts on Hamiltonian in \Eq{equationspace}, one shall have the following equation    
\beq 
\mathcal{P}H\left(\bm r,\bar{\bm \varepsilon}_r\left(\bm r \right),\bar{\bm \mu}_r\left(\bm r \right)\right)\mathcal{P}^{-1} = \sigma H\left(\bm -r,\bar{\bm \varepsilon}_r\left(-\bm r \right),\bar{\bm \mu}_r\left(-\bm r \right)\right){\sigma ^{ - 1}} = -H\left(\bm r,\bar{\bm \varepsilon}_r\left(-\bm r \right),\bar{\bm \mu}_r\left(-\bm r \right)\right).
\eeq 
If the cross-section of the waveguide is invariant under $\mathcal{P}$, i.e., $\bar{\bm \varepsilon}_r\left(\bm -r \right)=\bar{\bm \varepsilon}_r\left(\bm r \right)$ and $\bar{\bm \mu}_r\left(\bm -r \right)=\bar{\bm \mu}_r\left(\bm r \right)$, one obtains 
\beq\label{central symmetry}
\begin{split}
\mathcal{P}H\left(\bm r,\bar{\bm \varepsilon}_r\left(\bm r \right),\bar{\bm \mu}_r\left(\bm r \right)\right)\mathcal{P}^{-1}\mathcal{P} \Psi\left(\bm r\right) =-H&\left(\bm r,\bar{\bm \varepsilon}_r\left(\bm r \right), \bar{\bm \mu}_r\left(\bm r \right)\right) \mathcal{P} \Psi\left(\bm r\right)= \beta\sigma \Psi\left(-\bm r\right),\\
H\left(\bm r,\bar{\bm \varepsilon}_r\left(\bm r \right),\bar{\bm \mu}_r\left(\bm r \right)\right)\sigma\Psi\left(-\bm r\right)&=-\beta\sigma\Psi\left(-\bm r\right).
\end{split}
\eeq 
Consequently, $\mathcal{P}\Psi\left(\bm r\right)=\sigma\Psi\left(-\bm r\right)$ is the degenerated  mode (opposite propagation direction) of original state $\Psi\left(\bm r\right)$. In comparison  with \Eq{chiral symmetry} in chiral symmetry,  parity symmetry operation does not require that those components ($\bar {\bm \varepsilon} _r^{zt}$, $\bar {\bm \varepsilon} _r^{tz}$,$\bar {\bm \mu} _r^{zt}$ and $\bar {\bm \mu} _r^{tz}$) vanish, but reverses the coordinates of the fields before  performing  $\sigma$-operation.  Intuitively, it can be understood that the presence  of $\bar {\bm \varepsilon} _r^{tz}$ or $\bar {\bm \varepsilon} _r^{zt}$  elements in $\bar {\bm \varepsilon} _r$ or $\bar {\bm \mu} _r$ breaks the chiral symmetry between the forward and backward propagating modes, while the presence of parity symmetry in the structure of cross-section restore it. When $\bar{\bm\chi}$ can't be ignored, the conclusion will also be kept under $\bar{\bm \chi}\left(\bm r\right)=-\bar{\bm \chi}\left(-\bm r\right)$.

\subsection{$\mathcal{PT}$ symmetry}
In the time reversal/parity symmetry operation,   we have proved there is a definite relation between the forward and backward propagating modes that is guaranteed by  $\mathcal{P/T}$ symmetry. In this subsection, we continue to discuss the  the symmetric properties induced by combining the two symmetry operations together, i.e., $\mathcal{PT}$ symmetry, which has been examined extensively in the last few years \cite{Ganainy2007_OL,PT,ChongPRA2017,GePRX2014,XuOE2015}. As the operator $\mathcal{P}$ and $\mathcal{T}$ both act on Hamiltonian $H$, one obtains $\mathcal{P}\mathcal{T} H\left(\hat{p},\bm{r},t\right)\left(\mathcal{P}\mathcal{T}\right)^{-1}=H^*\left(\hat{p},-\bm{r},-t\right)$. If the optical systems are $\mathcal{PT}$ symmetric (here we only concerns isotropic medium), i.e., $\varepsilon_r\left(\bm r\right)= \varepsilon_r^*\left(-\bm r\right)$, $ \mu_r\left(\bm r\right)=\mu_r^*\left(-\bm r\right)$, one find the waveguide Hamiltonian $H$ commutes with the  $\mathcal{PT}$  operator, i.e., $\mathcal{PT} H\left(\mathcal{P}\mathcal{T}\right)^{-1}=H$, leading to 
\beq\label{PTsymmetry}
\mathcal{PT} H\left(\bm r,\bar{\bm \varepsilon}_r\left(\bm r\right),\bar{\bm \mu}_r\left(\bm r\right)\right)  \left(\mathcal{PT}\right)^{-1} \mathcal{PT}\Psi\left(\bm r\right) =H\left(\bm r,\bar{\bm \varepsilon}_r\left(\bm r\right),\bar{\bm \mu}_r\left(\bm r\right)\right) \Psi^*\left(-\bm r\right) = \beta^*\Psi^*\left(-\bm r\right).
\eeq
From \Eq{equationspace} and \Eq{PTsymmetry}, one immediately finds out the  fact  that if $\Psi\left(\bm r\right)$ is the eigenmode for Hamiltonian  with eigenvalue $\beta$, its complex conjugate partner with  reversed coordinates $\Psi^*\left(-\bm r\right)$ would also be the eigenmode with eigenvalue $\beta^*$. Before the $\mathcal{PT}$ symmetry is broken, the eigenvalues are always real number, with  $\beta^*=\beta$ and $\Psi^*\left(-\bm r\right)=\Psi\left(\bm r\right)$. Once the $\mathcal{
PT}$ symmetry is broken, $\beta^*$ and  $\beta$ are different values,  $\Psi^*\left(-\bm r\right)$ and $\Psi\left(\bm r\right)$ are separated eigenstates of $H$. The media can be anisotropy in time reversal symmetry or parity symmetry respectively, actually, the media under $\mathcal{PT}$ symmetry can also be anisotropy(See \Table{symmetrylist}).

\subsection{Rotation symmetry}
We continue to study the  degeneracy between the  polarization states $\Psi_1$, $\Psi_2$ due to the  rotational symmetry of the cross-section of waveguides. Considering the structure symmetry of the cross-section can be encoded into the optical properties of the material, we use \Eq{equationspace} that explicitly  encloses the coordinate-dependent material tensors, i.e., $\bar{\bm \varepsilon}_r\left(\bm r \right)$ and $\bar{\bm \mu}_r\left(\bm r \right)$. Due to the symmetry requirement,  we only consider isotropic waveguides such as ordinary optical fiber for simplicity. To this end,  the Hamiltonian $H$ can be reduced as,
\beq\label{hamspace}
H\left(\bm r,\bar{\bm \varepsilon}_r\left(\bm r \right),\bar{\bm \mu}_r\left(\bm r \right)\right)=\left( {\begin{array}{*{20}{c}}
  0&0&{ - {\partial _x}\frac{{{\partial _y}}}{{{k_0} \varepsilon _r}}}&{{\partial _x}\frac{{{\partial _x}}}{{{k_0} \varepsilon _r}} + {k_0} \mu _r} \\ 
  0&0&{ - {\partial _y}\frac{{{\partial _y}}}{{{k_0} \varepsilon _r}} - {k_0} \mu _r}&{{\partial _y}\frac{{{\partial _x}}}{{{k_0} \varepsilon _r}}} \\ 
  {{\partial _x}\frac{{{\partial _y}}}{{{k_0} \mu _r}}}&{ - {\partial _x}\frac{{{\partial _x}}}{{{k_0} \mu _r}} - {k_0} \varepsilon _r}&0&0 \\ 
  {{\partial _y}\frac{{{\partial _y}}}{{{k_0} \mu _r}} + {k_0} \varepsilon _r}&{ - {\partial _y}\frac{{{\partial _x}}}{{{k_0} \mu _r}}}&0&0 
\end{array}} \right).
\eeq
 If an eigenstates $\Psi_1\left(\bm r\right)$  in \Eq{equationspace} can be rotated anticlockwise by a constant angle $\theta$ to  another eigenstates $\Psi_2\left(\bm r\right)$, which can be described by the following equation,
\beq\label{twopola} 
\Psi_2\left(\bm r\right)=P\left(\theta\right)\Psi_1\left(R^{-1}\left(\theta\right)\bm r\right),
\eeq
where the polarization rotation operator
$ P_R\left(\theta \right)= \left( {\begin{array}{*{20}{c}}
  {\cos{\theta}}&{-\sin{\theta}}&0&0 \\ 
  {\sin{\theta}}&{\cos{\theta}}&0&0 \\ 
  0&0&{\cos{\theta}}&{-\sin{\theta}} \\
  0&0&{\sin{\theta}}&{\cos{\theta}}
  \end{array}} \right)$, and the coordinate rotation operator $
  R\left(\theta \right)= \left( {\begin{array}{*{20}{c}}
  {\cos{\theta}}&{-\sin{\theta}} \\ 
  {\sin{\theta}}&{\cos{\theta}} 
  \end{array}} \right)$. As can be seen, the rotation of the vector field is in sharp contrast to the rotation of a scalar field: if one wants to rotate a scalar anticlockwise, one just rotates the coordinate system clockwise by same angle;  as for vector field, one need to consider the rotation between the field components beside the rotation of each components, as described by \Eq{twopola}. As a side remark, $\Psi_1$ and $\Psi_2$ can be considered as  the polarization modes associated with the same field configuration, such that the in-plane vector fields of the two modes are always perpendicular, i.e.,  $\Psi_2=P_R\left(\frac{\pi}{2} \right)\Psi_1$. 
  
 As the operator $P_R(\theta)$  acts on the  Hamiltonian $H$, see \Eq{hamspace}, one shall obtain,
\beq\label{HR} 
\begin{split}
&P_R\left(\theta\right)H\left(\bm r,\bar{\bm \varepsilon}_r\left(\bm r \right),\bar{\bm \mu}_r\left(\bm r \right)\right){P_R^{ - 1}}\left(\theta\right)  = H\left(R\left(\theta\right)\bm r,\bar{\bm \varepsilon}_r\left(\bm r \right),\bar{\bm \mu}_r\left(\bm r \right)\right) \\
&=\left( {\begin{array}{*{20}{c}}
  0&0&{ - {\partial _u}\frac{{{\partial _v}}}{{{k_0} \varepsilon _r}}}&{{\partial _u}\frac{{{\partial _u}}}{{{k_0} \varepsilon _r}} + {k_0} \mu _r} \\ 
  0&0&{ - {\partial _v}\frac{{{\partial _v}}}{{{k_0} \varepsilon _r}} - {k_0} \mu _r}&{{\partial _v}\frac{{{\partial _u}}}{{{k_0} \varepsilon _r}}} \\ 
  {{\partial _u}\frac{{{\partial _v}}}{{{k_0} \mu _r}}}&{ - {\partial _u}\frac{{{\partial _u}}}{{{k_0} \mu _r}} - {k_0} \varepsilon _r}&0&0 \\ 
  {{\partial _v}\frac{{{\partial _v}}}{{{k_0} \mu _r}} + {k_0} \varepsilon _r}&{ - {\partial _v}\frac{{{\partial _u}}}{{{k_0} \mu _r}}}&0&0 
\end{array}} \right)
\end{split}
\eeq 
where  
$\left( {\begin{array}{*{20}{c}}
  u \\ 
  v \\ 
  \end{array}} \right)= \left( {\begin{array}{*{20}{c}}
  {\cos{\theta}}&{-\sin{\theta}} \\ 
  {\sin{\theta}}&{\cos{\theta}}\\ 
  \end{array}} \right)\left( {\begin{array}{*{20}{c}}
  x \\ 
  y \\ 
  \end{array}} \right)=R\left(\theta\right)\bm r$. It's interesting to note the fact that the operator $P_R(\theta)$ acting on $H$ is equivalent to rotate the differential Coordinates in $H$, with $\bar{\bm \varepsilon}_r\left(\bm r \right)$ and $\bar{\bm \mu}_r\left(\bm r \right)$ unchanged. With the substitution of  \Eq{twopola} and \Eq{HR} into \Eq{equationspace}, one obtains
\beq\label{rotation symmetry}
\begin{split}
P_R H\left(R^{-1}\bm r,\bar{\bm \varepsilon}_r\left(R^{-1}\bm r \right),\bar{\bm \mu}_r\left(R^{-1}\bm r \right)\right){P_R^{ - 1}}P_R\Psi_1\left(R^{-1}\bm r\right) &=\beta_1 P_R\Psi_1\left(R^{-1}\bm r\right), \\
H\left(\bm r,\bar{\bm \varepsilon}_r\left(R^{-1}\bm r \right),\bar{\bm \mu}_r\left(R^{-1}\bm r \right)\right)\Psi_2\left(\bm r\right) &=\beta_1 \Psi_2\left(\bm r\right).
\end{split}
\eeq
If the cross-section of waveguide is invariant under the rotation of $\theta$, we can get $H\left(\bm r,\bar{\bm \varepsilon}_r\left(\bm r \right),\bar{\bm \mu}_r\left(\bm r \right)\right)\Psi_2\left(\bm r\right)=\beta_1 \Psi_2\left(\bm r\right)$ from \Eq{rotation symmetry}, thus establishes the symmetric (degenerate) relation between the two polarization modes. When the media is on longer isotropy, we can get the same conclusion with constraint that $\hat{R}\bar{\bm \varepsilon}_r\left(R^{-1}\bm r\right)\hat{R}^{-1}=\bar{\bm \varepsilon}_r\left(\bm r\right)$, $\hat{R}\bar{\bm \mu}_r\left(R^{-1}\bm r\right)\hat{R}^{-1}=\bar{\bm \mu}_r\left(\bm r\right)$ and $R\bar{\bm \chi}\left(R^{-1}\bm r\right)R^{-1}=\bar{\bm \chi}\left(\bm r\right)$, where $\hat{R}=\left( {\begin{array}{*{20}{c}}
  {\cos{\theta}}&{-\sin{\theta}}&0 \\ 
  {\sin{\theta}}&{\cos{\theta}}&0\\ 
  0&0&1
  \end{array}} \right)$.

\begin{table}\centering
\setlength{\abovecaptionskip}{0.cm}
\setlength{\belowcaptionskip}{-0.cm}
\newcommand{\tabincell}[2]{\begin{tabular}{@{}#1@{}}#2\end{tabular}}
\caption{\label{symmetrylist} Symmetry properties of waveguide modes in the truncated mode set}
\begin{tabular}{c|c|c|c} \Xhline{1.5pt}
  &  Symmetry & Degeneracy & Constraints  \\ \hline\\
 \tabincell{c}{Chiral  \\symmetry}  &\tabincell{c}{
$\sigma H{\sigma ^{ - 1}} =  - H$}
& \tabincell{c}{$\beta^-=-\beta^+$\\
$\Psi^-=\sigma\Psi^+$} & \tabincell{c}{$\bar {\bm \varepsilon} _r^{zt}=\bar {\bm \varepsilon} _r^{tz}=0$\\ $\bar {\bm \mu} _r^{zt}=\bar {\bm \mu} _r^{tz}=0 $\\ and no $\bar{\bm \chi}$}\\
\hline
\tabincell{c}{Time reversal\\ symmetry}  &\tabincell{c}{$\mathcal{T} H{\mathcal{T}^{ - 1}} = -H$}
& \tabincell{c}{$\beta^-=-\left(\beta^+\right)^*$\\
$\Psi^-=\mathcal{T}\Psi^+=\sigma\left(\Psi^+\right)^*$}&  \tabincell{c}{$\bar{\bm \varepsilon}_r$, $\bar{\bm \mu}_r$ and $\bar{\bm \chi}$ are real }  \\
\hline
\tabincell{c}{Parity \\ symmetry}  &\tabincell{c}{
$ \mathcal{P}H\mathcal{P}^{-1}=-H$}& \tabincell{c}{$\beta^-=-\beta^+$\\ 
$\Psi^-\left(\bm r \right)=\mathcal{P}\Psi^+\left(\bm r \right)=\sigma\Psi^+\left(-\bm r \right)$}&  \tabincell{c}{ $\bar{\bm \varepsilon}_r\left(\bm r\right)=\bar{\bm \varepsilon}_r\left(-\bm r\right)$\\ $\bar{\bm \mu}_r\left(\bm r\right)=\bar{\bm \mu}_r\left(-\bm r\right)$\\ $\bar{\bm \chi}\left(\bm r\right)=-\bar{\bm \chi}\left(-\bm r\right)$}  \\ 
\hline
\tabincell{c}{$PT$ \\ symmetry}  &\tabincell{c}{
$\mathcal{PT} H (\mathcal{PT})^{-1}=H$}& \tabincell{c}{$\Psi_{\beta^*}\left(\bm r \right)=\mathcal{PT}\Psi_{\beta}\left(\bm r \right)=\Psi_{\beta}^*\left(-\bm r \right)$}&  \tabincell{c}{$\bar{\bm \varepsilon}_r\left(\bm r\right)= \bar{\bm \varepsilon}_r^*\left(-\bm r\right)$\\ $\bar{\bm \mu}_r\left(\bm r\right)=\bar{\bm \mu}_r^*\left(-\bm r\right)$\\ $\bar{\bm \chi}\left(\bm r\right)=-\bar{\bm \chi}^{*}\left(-\bm r\right)$  }\\
\hline
\tabincell{c}{Rotation  \\symmetry}  &\tabincell{c}{
$P_R H\left(R^{-1}\bm r\right){P_R^{ - 1}}= H\left(\bm r\right)$}& \tabincell{c}{$\beta_2=\beta_1$\\$\Psi_2\left(\bm r\right)=P_R\Psi_1\left(R^{-1}\bm r\right)$}&  \tabincell{c}{  $\hat{R}\bar{\bm \varepsilon}_r\left(R^{-1}\bm r\right)\hat{R}^{-1}=\bar{\bm \varepsilon}_r\left(\bm r\right)$\\ $\hat{R}\bar{\bm \mu}_r\left(R^{-1}\bm r\right)\hat{R}^{-1}=\bar{\bm \mu}_r\left(\bm r\right)$ \\$R\bar{\bm \chi}\left(R^{-1}\bm r\right)R^{-1}=\bar{\bm \chi}\left(\bm r\right)$ }  \\
\hline
\Xhline{1.5pt}
\end{tabular}
\end{table}
\subsection{Transformation of vector field }

To get a comprehensive impression  of  symmetry operations discussed in this paper, we list the five different  symmetry operations  in \Table{symmetrylist}. The first three symmetry operations are used to establish the symmetric relation between the forward and backward propagating modes, and the last two symmetry operations  establish the relationship between two  modes with same propagating direction.

The transverse electromagnetic field components, which are the eigenfunction $\Psi$ of waveguide Hamiltonian, is essentially a vector field. Considering the sharp contrast between rotating  vector fields  and rotating scalar fields, it is necessary to give  formal expressions to describe how the vector and scalar fields are rotated. According to \cite{scalar,scalarvector},  as  a rotating  operator $O_R$  acts on a scalar field (for example, $x$ component of electric field $e_x$) and a vector field (for example, transverse electric field ${\bm e}_t=\left(e_x,e_y\right)^T$), one shall have,
\beq\label{Rscalar}
O_R e_x\left(\bm r\right)=e_x\left(R^{-1}\bm r\right),
\eeq
and
\beq\label{Rvector}
O_R {\bm e}_t\left(\bm r\right)=\mathcal{R} {\bm e}_t\left(R^{-1}\bm r\right),
\eeq
where the operator $R$ is the aforementioned coordinate rotating operator, and $\mathcal{R}$ the rotating operation that reshuffles different components of the  vector fields. Rotating a scalar field is equivalent to rotating coordinates as described in \Eq{Rscalar}. Evident from \eqref{Rvector}, there are more evolved in the rotation of a vector field. In short, we could decompose the rotation of vector field into two steps: (1)  reshuffling the components of the vector field, and (2) coordinate rotation. Thus, the action of step (1) $\mathcal{R}{\bm e}_t\left(\bm r\right)$ and step (2) ${\bm e}_t\left(R^{-1}\bm r\right)$ are very different, one acting on the field components, the other on the coordinates of each components of the vector field.  We further explain the subtle difference via rotating the electric field, i.e., represented by the position-dependent arrows.  The $\mathcal{R}$  operator in \eqref{Rvector}   acts on the electric field directly (same as $\sigma$, $\mathcal{T}$ and $P_R$ in \Table{symmetrylist}),  only changes the orientation of the arrow  without moving the position of arrows, while the $R$ operation in ${\bm e}_t\left(R^{-1}\bm r\right)$   acts on the coordinates of the arrows, only changes the arrow position without  changing the orientation of the arrow. 

In Section 2.2 and 2.3, we only the reshuffle the components of the vector field without touching on the coordinates. Thus, the operator $\sigma$ in chiral symmetry and the operator  $\mathcal{T}$ in time reversal symmetry   essentially belongs to step (1). In Sections 2.4, 2.5 and 2.6, those symmetry  operations can be considered as combined operations of step (1) and step(2).
\begin{figure*}\includegraphics[width=1\textwidth]{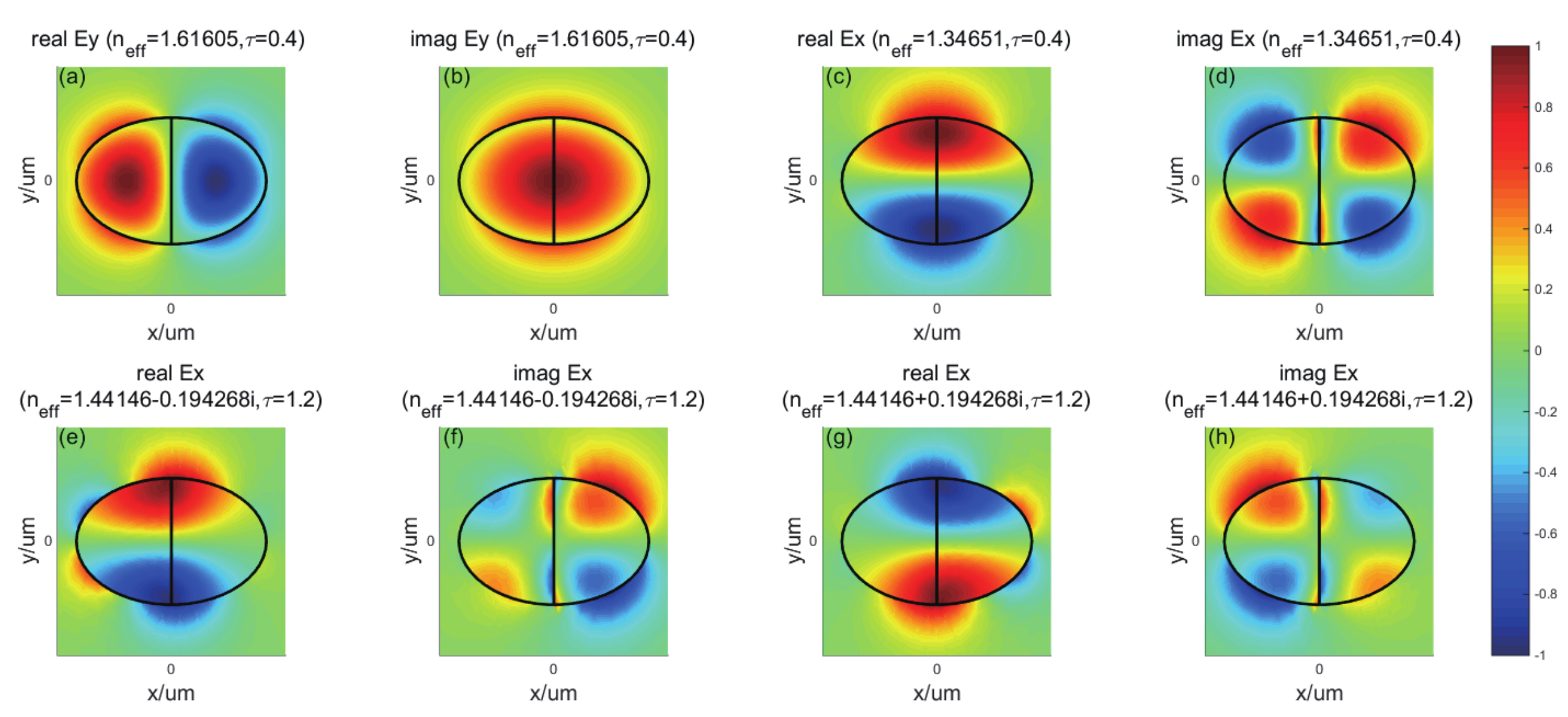}\caption{\label{figPT} The x and y-components of  the electric field for a pair of modes supported by the  gain-loss balanced waveguides before and after the exceptional point. The fields before/after EP are shown in the first/second row. (a/b) The dominating electric field component, i.e.,  $\textnormal{Re}(E_y)$/$\textnormal{Im}(E_y)$,  for  the mode with $n_{neff}=1.61605$;  (c/d) the dominating electric field component, i.e.,  $\textnormal{Re}(E_x)$/$\textnormal{Im}(E_x)$,  for  the mode with $n_{neff}=1.34651$. If  the $\mathcal{P}$ operation is applied, i.e., $\bm r\rightarrow -\bm r$, the real parts of the fields in (a,c) remains unchanged, while the imaginary parts of the fields in (b,d) change sign.  The x-components of the electric field, including $\textnormal{Re}(E_x)$ and $\textnormal{Im}(E_x)$,  for the two mode with conjugate $n_{eff}=1.44146 \pm 0.194268i$ are shown (e-h) . Evidently,  the field plot in (e) can be transformed to that in (g) under $\mathcal{P}$ operation. Similarly,  the field plot in (f) can be transformed to that in  (h) under $\mathcal{P}$ operation, but up to a sign difference.}
\end{figure*}
\section{Results and discussions}
\subsection{$\mathcal{PT}$ symmetry in gain-loss balanced waveguides}
The most commonly used optical structures in  $\mathcal{PT}$ symmetry systems are gain-loss balanced  waveguides. Here, we consider a simple example, see \Fig{figPT}, to illustrate the symmetric relations of the vector fields of a single mode or between two conjugated modes under the $\mathcal{PT}$ symmetry operation, depending on whether the $\mathcal{PT}$ symmetry breaking occurs or not. We consider elliptical  waveguide core with the semi-major (semi-minor) of $1.5\mu m$ ($1\mu m$). The  material in the waveguide core region is  isotropic, i.e., $\varepsilon_r=4-i\tau$ on the left hand side, while $\varepsilon_r=4+i\tau$ on the right hand side. The waveguide core is  embedded in air with operation  wavelength 4 $\mu$m. The eigen-fields and eigenvalues $\beta$  of the gain-loss balanced waveguides, as well as others throughout the paper are obtained by full-wave simulations  using COMSOL MULTIPHYSICS \cite{comsol}. As  the magnitude of gain/losses ($\tau$) increases, $\mathcal{PT}$ symmetry breaking occurs, the real parts of two  eigenvalues $\beta$ merger together and the overlapped imaginary part of the two  eigenvalues, i.e., $\textnormal{Im}(\beta)=0$,  bifurcates. The exact bifurcation  location of $\beta$ in $\tau$ is coined as the  exceptional point (EP).  As evident in \Fig{figPT}, the pair of modes with $n_{eff}=\beta/k_0$ of $1.61605$ and $1.34651$ ($\tau$=0.4)  evolve to the modes with $n_{eff}=1.44146 \pm 0.194268i$ ($\tau$=1.2)  as $\tau$ crosses EP (in-between 0.4 and 1.2). As shown in \Figs{figPT}  (a)-(d), the fields before EP   remains unchanged under   the subsequent    $\mathcal{P}$  ($\bm r\rightarrow -\bm r$)  and $\mathcal{T}$ (complex conjugation) operations. While both  the eigen-fields and eigenvalues $\beta$ of two modes after EP become  conjugate complex to each other,  see \Figs{figPT} (e)-(h), under the   subsequent $\mathcal{P}$ and $\mathcal{T}$  operations. Hence, the symmetric relations of the eigen-fields and eigen values shown \Fig{figPT} is consistent with  the predications by \Eq{PTsymmetry}.
  
It is worthy to point out that  the gain-loss balanced waveguide  also obeys chiral symmetry,  see discussion  in Section 2.2. Provided  one gets the eigen-field and eigenvalue $\beta_0$ of gain-loss balanced waveguides, as a consequence of chiral symmetry, -$\beta_0$ would also be the eigenvalue even  after EP. And the relationship between the modes with opposite eigenvalue is just given by the chiral operation. 

\begin{figure*}
\includegraphics[width=1\textwidth]{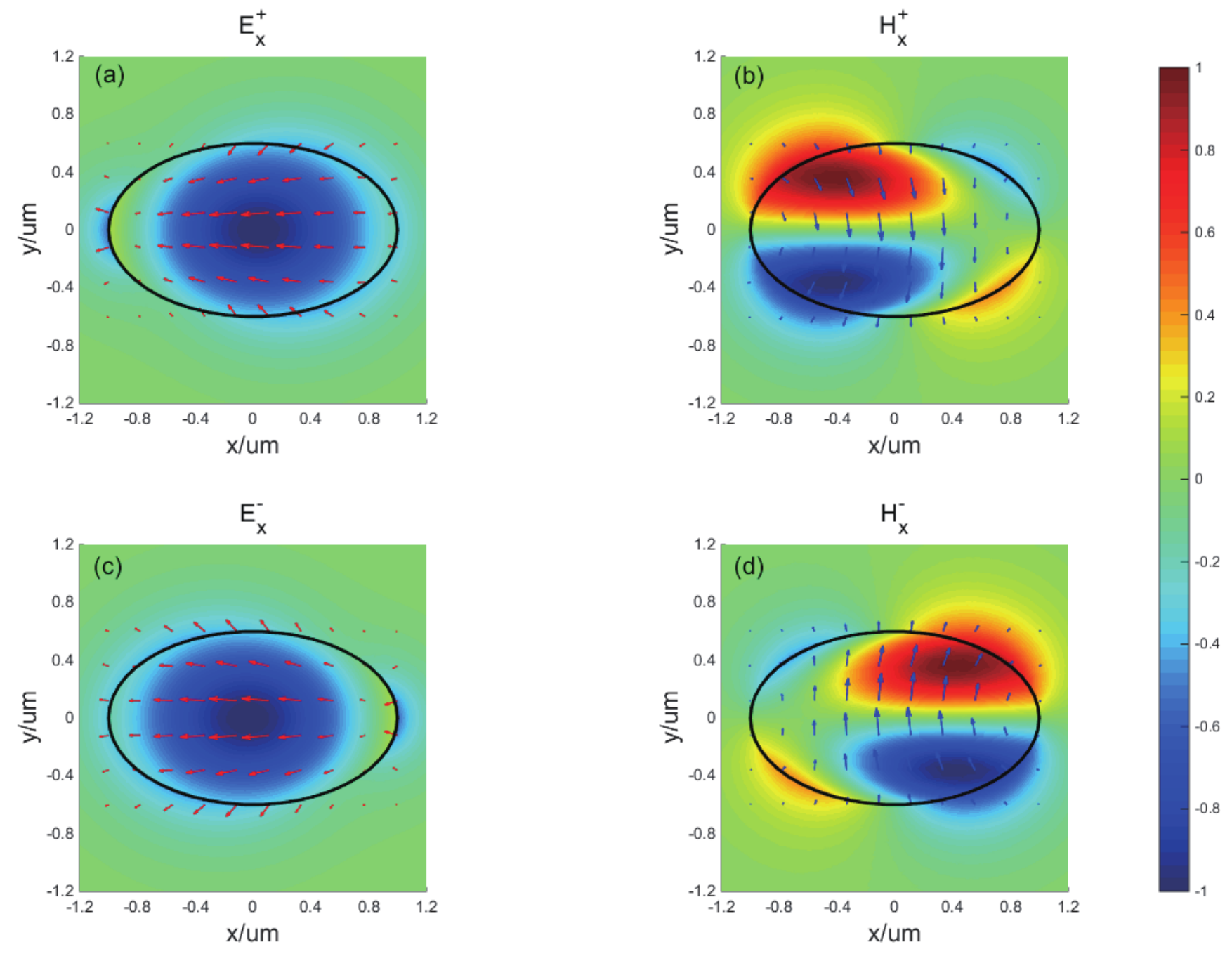}\caption{\label{fig2c} 
The $x$-component  as well as the vector field plots of the forward ((a)-(b)) and backward ((c)-(d)) propagating modes in anisotropic waveguide with ellipse-cross-section. The $x$-component of normalized electric/magnetic  field  is shown in (a,c)/(b,d), and  the vector plots of the in-plane electric field (magnetic field) are  also shown in  (a,c)/(b,d) indicated by the arrows, the length of which is proportional to the magnitude of the vector field. In (a) and (c), the vector field plots take  the real part of electric field, and the imaginary part of the magnetic field are taken in (b) and (d). Both  the effective refractive index of forward and backward modes  are $2.4668$.}
\end{figure*}
\subsection{Parity symmetry in anisotropic waveguides}
To illustrate the parity symmetry, we consider an anisotropic waveguide as shown in \Fig{fig2c}. The cross-section of the waveguide is elliptical, thus has the $C_{2z}$ symmetry, which is equivalent to the coordination transform $\bm r\rightarrow -\bm r$ in the 2D transverse plane. In the waveguide, the semi-major and semi-minor axis are 1 $\mu$m and 0.6 $\mu$m, respectively. The relative permittivity  is $\bar{\bm \varepsilon}_r=\begin{pmatrix}
10&0&4i \\
0&10&0 \\
-4i&0&10 \\ 
\end{pmatrix}$ corresponding to magneto-optical materials, and the permeability $\mu_r$  1, with background medium  air. As predicated  in Section 2.4, the transverse electric fields in backward  propagating mode are same as that  in forward mode under the parity operation ($\bm r\rightarrow -\bm r$), while the magnetic field transforms in a similar fashion but acquire an additional sign flip. Comparing \Fig{fig2c} (a-b) with (c-d), it's clear that the electric field $x$ component and magnetic field $x$ component are consistent with the predications from Section 2.4. 
\begin{figure*}
\includegraphics[width=1\textwidth]{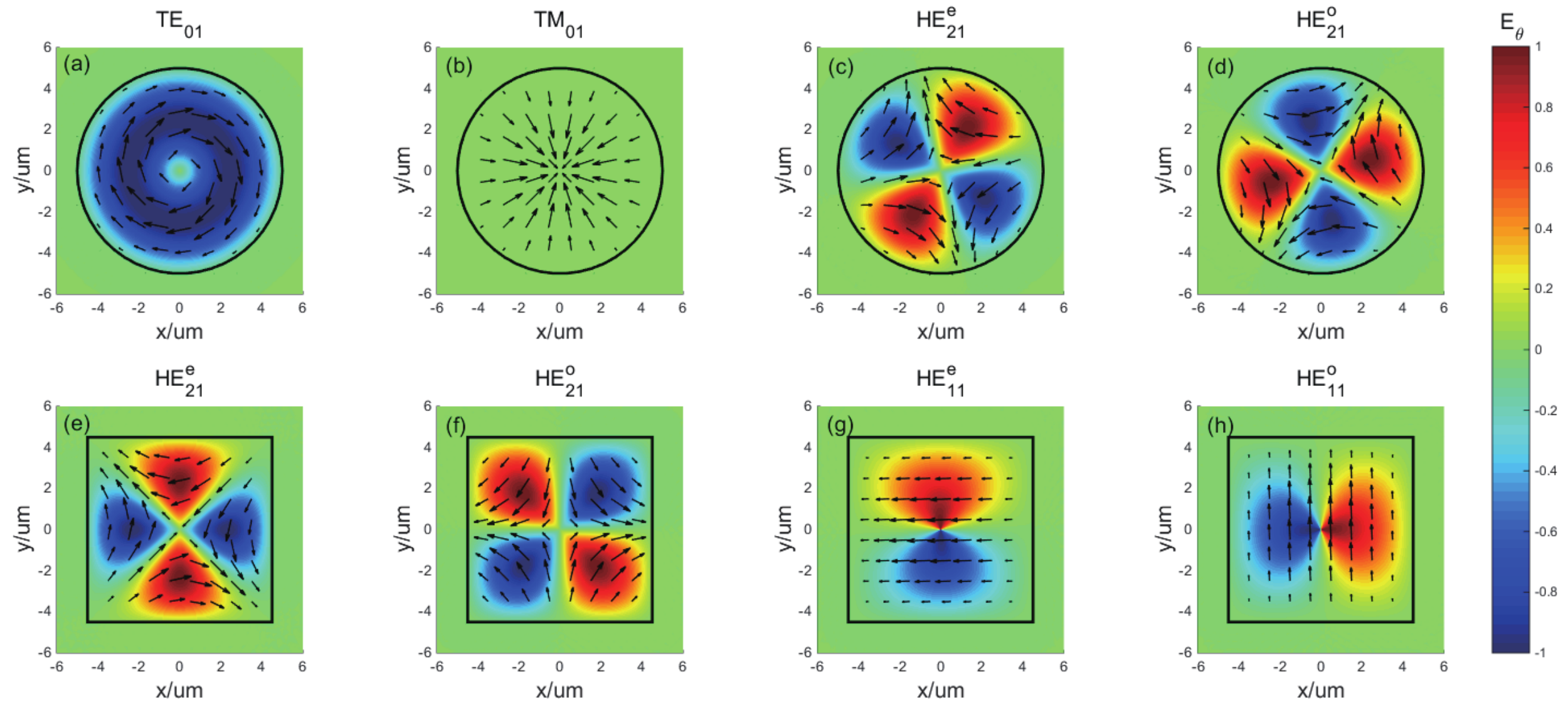}\caption{\label{fig2} (a-d) TE$_{01}$, TM$_{01}$, HE$_{21}^e$ and HE$_{21}^o$ in circular  optical fibers. The four modes reduce into LP$_{11}$ under weakly guiding. (e-h) HE$_{21}^e$, HE$_{21}^o$, HE$_{11}^e$ and HE$_{11}^o$ in square-core waveguide. The black circle or square indicate the borders of the waveguide-core, and arrows show the electric field orientation. The color plots are the azimuthal component, i.e., $E_{\theta}$, of  electric field. The radius of circle-core in (a-d) is 5 $\mu$m, the relative permittivity $\varepsilon_r$ is 4, and  the background material is air. The effective refractive index is $1.99154$ in (a), $1.99117$ in (b), and $1.99135$ in both (c) and (d). The  length of square in (e-h) is $9\mu m$,  the material properties are the same as (a-d). The effective refractive index is $1.99118$ in (e), $1.99093$ in (f), and $1.99643$ in both (g) and (h).  Work wavelength is $1.55\mu m$  }
\end{figure*}
\subsection{Rotational symmetry in optical fiber}
According to the dual symmetry of Maxwell equation, the two forward propagating modes  $\Psi_1$ and $\Psi_2$ are degenerated provided $\varepsilon_r= \mu_r$, which  can be easily proved by exchanging the permittivity tensor and permeability tensor in the Hamiltonian. In the following, we will show that rotational symmetries in optical waveguides can protect the degeneracy of the two forward propagating modes $\Psi_1$ and $\Psi_2$ without $\varepsilon_r=\mu_r$ via concrete examples, i.e., circular optical fiber or square optical waveguides, under certain conditions. 

This differences between the  pure TE/TM modes  and the   HE/HE modes lead  to the following fact: one  mode in  each   HE/EH mode pair within the aforementioned truncated mode set in circular fiber can be transformed to the other by rotating their transverse fields globally with a constant angle,  such statement does not hold for  pure TE/TM modes, see details in Apppendix A. As an example, we pick out four modes of optical fiber as shown in \Fig{fig2} (a-d). Two observations can be seen: (1)  despite the variation of the rotational angle, the TE$_{01}$  in  \Fig{fig2} (a)  or TM$_{01}$  in \Fig{fig2} (b) can only be rotated to itself  \cite{RichardBook2010};  (2) while the field orientations of HE$_{21}^e$ (\Fig{fig2}(c)) and HE$_{21}^o$ (\Fig{fig2}(d)) can be exchanged by rotating $\pi/4$. In consistency with the discussion in Section 2.7, the operator $P_R\left(\theta\right)\Psi\left(R\left(\theta\right)^{-1}\bm r\right)$ can be seen as a global  rotation of  the electric field orientation in \Fig{fig2} with angle $\theta$.  Therefore, the HE$_{21}^e$ and HE$_{21}^o$  satisfy \Eq{twopola}  with the angle  $\theta=\frac{\pi}{4}$. Moreover, for any HE/EH mode  pair satisfies  \Eq{twopola} with corresponding rotating angle, i.e., $\frac{\pi}{2}$ for HE$_{11}$ ($l=0$) and $\frac{\pi}{6}$ for HE$_{31}$ ($l=2$), the polarization degeneracy exists. Due to the fact that  the circle-core fiber has continuous rotational symmetry, any  HE/EH mode pair are degenerated.  As for  TE$_{01}$ and TM$_{01}$, there is no such relation for any mode pair.   As for a waveguide with discrete rotational symmetry, if  the azimuthal quantum number ($l$) of modes are consistent with the discrete symmetry of waveguide core, the degeneracy emerges. Otherwise the degeneracy vanishes. For example, in square-core fiber, the HE$_{21}$ shown in \Figs{fig2} (e)-(f) modes are not degenerate, because square is not invariant under rotating with $\frac{\pi}{4}$. However, the HE$_{11}$ (\Fig{fig2}(g-h)) modes in square-core fiber is also degenerate, since square cross section remains the same under rotation of $\frac{\pi}{2}$.

\section{Conclusion}
In conclusion, we provide a systematic approach to classify the symmetric properties  of waveguide modes  in presence of gain/losses, anisotropy/bi-anisotropy, as well as the rational symmetry in the geometric cross-section. By eliminating the longitudinal field components ($e_z$ and $h_z$), we derive the waveguide Hamiltonian that fully characterizes the waveguide modes. With the proper symmetry operations, i.e., chiral/time reverse/ parity symmetry, associated with waveguide Hamiltonian, one can easily build up the relations between  forward and backward propagating modes. As for the $\mathcal{PT}$ symmetry, we can identity the symmetric properties of mode profile if the $\mathcal{PT}$ symmetry is fulfilled. For the cross-section with rotational symmetry, we study how the rotation symmetry gives rise to the polarization degeneracy, illustrated by circular  fiber and  waveguide with square cross-section.

The derived Hamiltonian as well as the symmetry operation don't rely on any specific material parameters or chosen geometry, thus can be applied  to any waveguide system, as long as certain symmetry relation (not limited within the symmetries discussed in the paper) is fulfilled . Importantly, our approach can be applied to analysis waveguide modes without knowing the exact field distribution, thus simplifies the modal analysis and can be useful for wave-guiding design.

\section*{Appendix A: Rotational symmetry of vector field in circular fiber}
In circular fiber with strong guidiance, the waveguide modes are described by TE$_{0m}$ (transverse electric modes), TM$_{0m}$ (transverse magnetic modes), and HE$_{lm}^p$ and EH$_{lm}^p$ (the last two are hybrid modes), depending on the existence and weighting of $E_z$ and  $H_z$. The indices $l$s represent the azimuthal quantum numbers, and $m$s radial quantum numbers. The pure TE and TM modes are  special in the sense that all the field components have  no azimuthal dependence in cylindrical coordinate $\left(\rho,\phi,z\right)$,  such as TE$_{01}$ in \Fig{fig2} (a).  We can reformulate Hamiltonian in polar coordinate,
\beq
H\left(\rho,\phi\right)=\left( {\begin{array}{*{20}{c}}
  0&0&{ - {\partial _\rho}\frac{{{\partial _\varphi}}}{{{k_0} \varepsilon _r\rho}}}&{{\partial _\rho}\frac{{{\partial _\rho}}}{{{k_0} \varepsilon _r\rho}}\rho + {k_0} \mu _r} \\ 
  0&0&{ - {\frac{\partial _\phi}{\rho}}\frac{{{\partial _\phi}}}{{{k_0} \varepsilon _r\rho}} - {k_0} \mu _r}&{{\frac{\partial _\phi}{\rho}}\frac{{{\partial _\rho}}}{{{k_0} \varepsilon _r\rho}}}\rho \\ 
  {{\partial _\rho}\frac{{{\partial _\phi}}}{{{k_0} \mu _r\rho}}}&{ - {\partial _\rho}\frac{{{\partial _\rho}}}{{{k_0} \mu _r\rho}}\rho - {k_0} \varepsilon _r}&0&0 \\ 
  {{\frac{\partial _\phi}{\rho}}\frac{{{\partial _\phi}}}{{{k_0} \mu _r\phi}} + {k_0} \varepsilon _r}&{ - {\frac{\partial _\phi}{\rho}}\frac{{{\partial _\rho}}}{{{k_0} \mu _r\rho}}}\rho&0&0 
\end{array}} \right),
\eeq
with eigenstate $\Psi=\left[e_{\rho}, e_{\phi},h_{\rho},h_{\phi}\right]^T$.  For TE and TM modes, $\partial _\phi=0$, thus $H\left(\rho,\phi\right)$ reduces to the form that only has anti-diagonal elements,  leading to the decoupling of TE and TM modes with different eigenvalues. Thus, the TE and TM modes in circular fiber can never be degenerated, though they share the same quantum number (They are approximately degenerate with weak guidiance).

Due the continuous rotational symmetry, all fields in optical fiber can be conveniently expressed as a $\rho$-function multiplied by a $\phi$-function, where  $\rho$ and $\phi$  are the radial and azimuthal variables \cite{opticalwaveguide}. For a given TE/TM, or EH/HE mode labeled by the quantum number $(m,l)$ , the transverse electric (magnetic) field can be written as 
\beq\label{eq1}
\bm{e}_t^{e/ o}=J_m^1\left(\rho\right)\begin{pmatrix}
 \cos\left(l\phi\right)\\
\sin\left(l\phi\right) \\
\end{pmatrix} or \begin{pmatrix}
 \sin\left(l\phi\right)\\
-\cos\left(l\phi\right) \\
\end{pmatrix},
\eeq
or 
\beq\label{eq2}
\bm{e}_t^{e/ o}=J_m^2\left(\rho\right)\begin{pmatrix}
 \sin\left(l\phi\right)\\
\cos\left(l\phi\right) \\
\end{pmatrix} or \begin{pmatrix}
 -\cos\left(l\phi\right)\\
\sin\left(l\phi\right) \\
\end{pmatrix}.
\eeq 
Evidently, in cylindrical coordinates, \Eq{eq1} represents EH$_{(l-1)m}$ modes when $l>1$ or TE$_{0m}$ and TM$_{0m}$ modes when $l=1$, while \Eq{eq2} for HE$_{(l+1)m}$ modes. If $l>1$ and $\left(l-1\right)\theta=\frac{\pi}{2}$, we can have  the field in  \Eq{eq1} under the rotation by \Eq{twopola} given by 
\beq
\begin{pmatrix}
 \cos\left(\theta\right)&-\sin\left(\theta\right)\\
\sin\left(\theta\right)&\cos\left(\theta\right) \\
\end{pmatrix}\begin{pmatrix}
 \cos\left(l\left(\phi-\theta\right)\right)\\
\sin\left(l\left(\phi-\theta\right)\right) \\
\end{pmatrix}=\begin{pmatrix}
 \cos\left(l\phi-\left(l-1\right)\theta\right)\\
\sin\left(l\phi-\left(l-1\right)\theta\right) 
\end{pmatrix}=\begin{pmatrix}
 \sin\left(l\phi\right)\\
-\cos\left(l\phi\right) \\
\end{pmatrix},
\eeq
which means that for pair of EH modes, we can rotate one vector field with $\theta=\frac{\pi}{2\left(l-1\right)}$ to get another one. However, it's invalid for TE and TM modes due to  $l=1$. Similarly in \Eq{eq2},  we have,
\beq
\begin{pmatrix}
 \cos\left(\theta\right)&-\sin\left(\theta\right)\\
\sin\left(\theta\right)&\cos\left(\theta\right) \\
\end{pmatrix}\begin{pmatrix}
 \sin\left(l\left(\phi-\theta\right)\right)\\
\cos\left(l\left(\phi-\theta\right)\right) \\
\end{pmatrix}=\begin{pmatrix}
 \sin\left(l\phi-\left(l+1\right)\theta\right)\\
\cos\left(l\phi-\left(l+1\right)\theta\right) 
\end{pmatrix}=\begin{pmatrix}
 -\cos\left(l\phi\right)\\
\sin\left(l\phi\right) \\
\end{pmatrix},
\eeq
provided $\left(l+1\right)\theta=\frac{\pi}{2}$. This implies that  for HE modes, we can rotate one vector field with $\theta=\frac{\pi}{2\left(l+1\right)}$ to get another one.

\section*{Acknowledgment}
Y. Chen acknowledges financial support from the National Natural Science Foundation of China (Grant No. 61405067), and  the Fundamental Research Funds for the Central Universities, HUST: 2017KFYXJJ027. 
\end{document}